\documentstyle[12pt,world_sci]{article}

\newcommand{\be}{\begin{equation}}
\newcommand{\ee}{\end{equation}}
\newcommand{\beq}{\begin{eqalignno}}
\newcommand{\eeq}{\end{eqalignno}}
\newcommand{\epem}{\mbox{$e^+e^-$}}

\newcommand{\gev}{{\rm\,GeV}}

\newcommand{\fb}{{\rm\,fb}}

\newcommand{\ifb}{{\rm\,fb}^{-1}}

\newcommand{\mw}{M_W}
\newcommand{\cosb}{\cos\beta}
\newcommand{\sinb}{\sin\beta}
\newcommand{\tanb}{\tan\beta}

\newcommand{\chc}{\tilde{\chi}^{\pm}}
\newcommand{\chcflip}{\tilde{\chi}^{\mp}}
\newcommand{\chargino}{\tilde{\chi}^{\pm}_1}
\newcommand{\charginotwo}{\tilde{\chi}^{\pm}_2}
\newcommand{\mchargino}{m_{\tilde{\chi}^{\pm}_1}}
\newcommand{\mcharginotwo}{m_{\tilde{\chi}^{\pm}_2}}
\newcommand{\chcp}{\tilde{\chi}^+}
\newcommand{\chcm}{\tilde{\chi}^-}
\newcommand{\chn}{\tilde{\chi}^0}
\newcommand{\LSP}{\chn_1}
\newcommand{\mLSP}{m_{\chn_1}}
\newcommand{\snu}{\tilde{\nu}}
\newcommand{\msnu}{m_{\tilde{\nu}}}

\newcommand{\mslep}{m_{\tilde{l}}}

\newcommand{\msq}{m_{\tilde{q}}}

\newcommand{\mum}{(\mu , M_2)}

\newcommand{\ds}{\displaystyle}

\newcommand{\mwchi}{M_W^{\chi}}
\newcommand{\gchi}{g^{\chi}}
\newcommand{\dsigr}{d\sigma_R/d\cos\theta}
\newcommand{\dsigl}{d\sigma_L/d\cos\theta}

\newcommand{\afbtrunc}{A_{FB}^{\chi '}}

\newcommand{\third}{\frac{1}{3}}

\newcommand{\rarr}{\rightarrow}

\begin{document}
\setlength{\baselineskip}{3.9ex}

\begin{flushright}
SLAC--PUB--6662 \\
hep-ph/9409264 \\
September 1994 \\
(M) \\
\end{flushright}

\title{{\bf TESTING SUPERSYMMETRY AT THE NEXT LINEAR COLLIDER}}
\bigskip
\author{
JONATHAN L. FENG
\thanks{Work supported by the Department of Energy, contract
DE--AC03--76SF00515, and in part by an NSF Graduate Research
Fellowship.}
\thanks{E-mail address: jlf@slacvm.slac.stanford.edu.}
\\
\medskip
{\em Stanford Linear Accelerator Center \\
Stanford University, Stanford, California 94309} \\
\medskip
and \\
\medskip
{\em Physics Department \\
Stanford University, Stanford, California 94305}
}

\maketitle
\setlength{\baselineskip}{3.9ex}

\begin{center}
\parbox{13.0cm}
{\begin{center} ABSTRACT \end{center}
{\small \hspace*{0.3cm}
\setlength{\baselineskip}{3.9ex}
If new particles are discovered, it will be important to determine if
they are the supersymmetric partners of standard model bosons and
fermions. Supersymmetry predicts relations among the couplings and
masses of these particles.  We discuss the prospects for testing these
relations at a future $e^+e^-$ linear collider with measurements that
exploit the availability of polarized beams.

\vspace*{0.7in}

\begin{center}
Talk presented at \\
DPF'94: Eighth Meeting of the Division of Particles and Fields \\
of the American Physical Society \\
Albuquerque, New Mexico, August 2--6, 1994
\end{center}
}}
\end{center}

\newpage
\setlength{\baselineskip}{3.9ex}
\pagestyle{plain}

\section{Introduction}
\label{sec:Introduction}

The phenomenological predictions of supersymmetry (SUSY) may be divided
into the following three categories: (I) predictions that would
constitute indirect evidence for SUSY if verified, including, for
example, the existence of a light Higgs boson; (II) the existence of
particles with the correct spin and quantum numbers to be superpartners
of standard model particles; and (III) well-defined quantitative
relations among the couplings and masses of these new particles. While
the predictions of (I) are of great interest, their verification is
clearly no substitute for direct evidence.  The discovery of a large
number of particles in category (II) would be strong support for SUSY,
but it is unlikely that future searches will immediately yield many
sparticles.  On the other hand, if only one or a few new particles are
discovered, precise verification of the relations of category (III)
could be taken as confirmation of SUSY. It is the prospects for such
tests that we investigate in this study.

Studies have shown that the Next Linear Collider, a proposed linear
$\epem$ collider with $\sqrt{s} = 500 \gev$ and a luminosity of
$10-100\ifb$/year, is a powerful tool for studying the properties of
new particles.\cite{JLC} The clean environment and polarizable beams
provided by such a machine make it well-suited to precision studies,
and we will study the prospects for precision tests of SUSY in this
setting. We will also limit the discussion to the case in which
charginos are produced, but slepton and squark pair production is
beyond reach. Remarks about other scenarios may be found in an extended
version of this work done with H.~Murayama, M.~E.~Peskin, and
X.~Tata.\cite{longpaper}

In Sec.~\ref{sec:Regions} we assume SUSY and use it as a guide to
selecting well-motivated case studies.  We review the parameters that
enter chargino production and decay, and we divide the SUSY parameter
space into characteristic regions. In Sec.~\ref{sec:Region 1} we
explore the first of these regions and test the form of the chargino
mass matrix. In Sec.~\ref{sec:Region 2} we consider another region and
test the chargino-fermion-sfermion coupling.

\section{Regions of Parameter Space}
\label{sec:Regions}

This study will be conducted in the context of the minimal
supersymmetric standard model (MSSM), the supersymmetric extension of
the standard model with minimal field content. The charginos of the
MSSM are mixtures of the charged Higgsinos and electroweak gauginos and
have mass terms $(\psi ^-)^T {\bf M}_{\chc} \psi^+ + {\rm h.c.}$, where
\be\label{chamass}
{\bf M}_{\chc} = \left( \begin{array}{cc}
 M_2                    &\sqrt{2} \, \mw\sinb  \\
\sqrt{2} \, \mw\cosb   &\mu                    \end{array} \right),
\ee
and $(\psi^{\pm})^T = (-i\tilde{W}^{\pm}, \tilde{H}^{\pm})$. The
chargino mass eigenstates are $\tilde{\chi}^+_i = {\bf V}_{ij}\psi^+_j$
and $\tilde{\chi}^-_i = {\bf U}_{ij}\psi^-_j$. The matrices $\bf V$ and
$\bf U$ are effectively orthogonal rotation matrices parametrized by
the angles $\phi_+$ and $\phi_-$, respectively.

We assume that R-parity is conserved, the LSP is the lightest
neutralino $\LSP$, there is no intergenerational mixing in the sfermion
sector, and sleptons and squarks are degenerate with masses $\mslep$
and $\msq$, respectively. (The last assumption may be partially
relaxed.\cite{LEPIIstudy}) With these assumptions, the parameters that
enter chargino events are $\mu$, $M_2$, $\tanb$, $M_1$, $\mslep$, and
$\msq$. With an $e^-_L$ beam, chargino production occurs through
$s$-channel $Z$ and $\gamma$ diagrams and $t$-channel $\snu_e$
exchange, and so $\dsigl$ is governed by the first four parameters. In
the case of an $e^-_R$ beam, the $\snu_e$ diagram is absent, and so
$\dsigr$ is dependent on only the first three parameters $\mu$, $M_2$,
and $\tanb$. Charginos decay to the LSP either leptonically through $W$
bosons or virtual sleptons, $\chcp\rightarrow (\LSP W^+, {\tilde{l}}^*
\nu, \bar{l} {\tilde{\nu}}^*)\rightarrow \LSP \bar{l} \nu$, or
hadronically through $W$ bosons or virtual squarks, $\chcp\rightarrow
(\LSP W^+, {\tilde{q}}^* q', \bar{q} {\tilde{q'}}^*) \rightarrow \LSP
\bar{q} q'$. All six parameters enter the decay process.

We now divide the parameter space into characteristic regions.  The
chargino masses and $\sigma_R \equiv \sigma(e^-_R e^+_L \rarr \chcp_1
\chcm_1)$ depend only on $\mu$, $M_2$, and $\tan\beta$.  The dependence
on $\tan\beta$ is weak; we set $\tan\beta=4$ as a representative case.
In Fig. 1, the cross-hatched region is excluded by present experiments,
and chargino production is inaccessible for $\sqrt{s} = 500\gev$ in the
hatched region.  We divide the remaining bands of the $\mum$ plane into
the three regions indicated. In region 1, $\chargino\chcflip_2$
production is possible, and so both chargino masses can be measured.
Where $\charginotwo$ is inaccessible, $\sigma_R$ distinguishes regions
2 (shaded, $\sigma_R\approx 0$) and 3 (where, typically,  $\sigma_R> 50
\fb$). We will consider case studies in which $\mchargino\approx 170
\gev$; this contour is the dotted curve in Fig.~1. Region 3 presents
difficulties even for the identification of a SUSY signal, since the
$\chargino$ and $\LSP$ become degenerate as $M_2$  increases. Although
it may be possible to verify SUSY relations in certain parts of region
3, we will not consider this case further.

\section{Region 1}
\label{sec:Region 1}

In region 1 we take the representative point in parameter space to be
($\mu$, $M_2$, $\tanb$, $M_1/M_2$, $\mslep$, $\msq$) = ($-195$, 210, 4,
0.5, 400, 700). For these parameters, $\mchargino = 172 \gev$, $\mLSP =
105 \gev$, and $\mcharginotwo = 255 \gev$. The uncertainty in
determining these masses is very small \cite{JLC} and will be
unimportant for this study. There are additional features that are
typical of Region 1: $\sigma_R$ is large enough to yield many events
for study, and $\chargino \approx \tilde{H}^\pm$, so the leptonic
branching fraction $B_l$ is $\third$ to a very good approximation.

In this region we generalize the chargino mass matrix to an arbitrary
real $2\times 2$ matrix, which we parametrize as
\be\label{chamasschi}
{\bf M}_{\chc} = \left( \begin{array}{cc}
 M_2                    &\sqrt{2} \, \mwchi\sinb  \\
\sqrt{2} \, \mwchi\cosb   &\mu                    \end{array}
\right).
\ee
Our goal is to test the SUSY relation $\mwchi = \mw$, that is, the
equality of the Higgs boson and Higgsino couplings. Formally, this is a
simple task.  The four parameters entering Eq.~\ref{chamasschi} may be
exchanged for the parameters $\mchargino$, $\mcharginotwo$, $\phi_+$
and $\phi_-$.  By measuring $\mchargino$, $\mcharginotwo$, and two
quantities derived from $\dsigr$, we may restrict the variables
($\phi_+$, $\phi_-$) and may therefore bound $\mwchi$.  It is useful
to work with the total cross section $\sigma_R$ and a truncated
forward-backward asymmetry $\afbtrunc$, defined below. Unfortunately,
neither quantity is observed directly.  We have performed Monte Carlo
simulations at a large number of points obtained by varying the
supersymmetry parameters and $\mwchi$ to determine the correlation of
these quantities to experimental observables.

The Monte Carlo simulations for chargino events used the parton-level
event generator of Feng and Strassler \cite{LEPIIstudy} with detector
parameters as chosen in the JLC study.\cite{JLC}  Chargino events were
selected from mixed-mode events in which one chargino decays
hadronically and the other leptonically, using the system of cuts
presented by the JLC group.  These cuts include the elimination of
events with very forward leptons or hadron jets ($\cos\theta_{\rm had},
\cos\theta_l < 0.75$), to remove the forward peak of $WW$ events. With
these cuts and a highly polarized $e^-_R$ beam, the total background is
negligible.

To find $\sigma_R$, we use the event rate in the mixed mode, the
leptonic branching ratio $B_l = \third$, and the efficiency of the cuts
determined by Monte Carlo.  The simulations also tell us that the
theoretical quantity
\be
\afbtrunc \equiv \frac
{\ds \sigma^\chi (0<\cos\theta<0.75)-\sigma^\chi (-1<\cos\theta<0)}
{\ds \sigma^\chi (-1<\cos\theta<0.75)}
\ee
is highly correlated with the forward-backward asymmetry of the
hadronic system's direction.  For an integrated luminosity of $50\ifb$,
these two quantities should be determined to the level $\sigma_R = 48
\pm 2.4$ fb, $\afbtrunc  = -37\pm 6.9\%$, where the 1$\sigma$ errors
include uncertainty from the variation of parameters. The measurements
of $\afbtrunc$ and $\sigma_R$ constrain the $(\phi_+, \phi_-)$ plane to
the shaded region in Fig.~2.  In this allowed region, $65\gev < \mwchi
< 100\gev$, a significant quantitative confirmation of SUSY.

\section{Region 2}
\label{sec:Region 2}

In region 2 we take the representative point to be $(\mu, M_2, \tanb,
M_1/M_2, \mslep, \msq) = (-500, 170, 4, 0.5, 400, 700)$. For these
parameters, $\mchargino = 172 \gev$, $\mLSP = 86 \gev$, $\mcharginotwo
= 512 \gev$, and $\sigma_R \approx 0$. Here we must rely on measurement
of $\dsigl$, which introduces dependence on $\msnu$. Fortunately, there
is a compensating simplification: in region 2, $\phi_+,\phi_- \approx
0$, i.e., charginos and neutralinos are very nearly pure gauginos.  In
addition, as is typical in region 2, on-shell $W$ decays are allowed,
and so again $B_l = \third$.

We will generalize the $\chargino f \tilde{f}$ coupling to $\gchi {\bf
V}_{11}$ and test the SUSY relation $\gchi=g$, that is the equality of
the $W$ boson and wino couplings. The differential cross section
$\dsigl$ is a function of $(\mchargino, \phi_+, \phi_-, \msnu, \gchi)$,
but because we can measure $\mchargino$, and $\phi_+, \phi_- \approx
0$, we have only two unknowns.  These may be constrained with two
quantities formed from $\dsigl$, in particular, $\sigma_L$ and
$\afbtrunc$.

We follow the procedure of the previous section, with the exception of
using cuts appropriate to on-shell $W$ decays.\cite{Grvz2} Including
all errors, we find that $\afbtrunc = 20 \pm 5.3\%$ and
$\Delta\sigma_L/\sigma_L = 6.4\%$ for an integrated luminosity of
$50\ifb$.  These measurements constrain the allowed region of the
$(\msnu,\gchi)$ plane to the three shaded areas shown in Fig.~3. If
$\msnu< 250 \gev$ can be excluded, the allowed region is only the
largest of these shaded regions, in which $0.75g \le \gchi \le 1.3g$.
In addition to confirming the prediction of SUSY, it is clear from
Fig.~3 that we have simultaneously bounded $\msnu$, a useful result for
future sparticle searches.

\def\cmp#1{{\it Comm. Math. Phys.} {\bf #1}}
\def\pl#1{{\it Phys. Lett.} {\bf #1B}}
\def\prl#1{{\it Phys. Rev. Lett.} {\bf #1}}
\def\prd#1{{\it Phys. Rev.} {\bf D#1}}
\def\prr#1{{\it Phys. Rev.} {\bf #1}}
\def\prb#1{{\it Phys. Rev.} {\bf B#1}}
\def\np#1{{\it Nucl. Phys.} {\bf B#1}}
\def\ncim#1{{\it Nuovo Cimento} {\bf #1}}
\def\jmp#1{{\it J. Math. Phys.} {\bf #1}}
\def\mpl#1{{\it Mod. Phys. Lett.} {\bf A#1}}

\bibliographystyle{unsrt}

\begin{thebibliography}{99.}

\bibitem{JLC}
JLC Group, {\it JLC--I}, Tsukuba, Japan, 1992 (KEK Report 92--16,
Tsukuba, 1992).

\bibitem{longpaper}
J.L. Feng, H. Murayama, M.E. Peskin, and X. Tata, SLAC--PUB--6654,
LBL--36101, UH--511--804--94 (1994).

\bibitem{LEPIIstudy}
J.L. Feng and M. Strassler, SLAC--PUB--6497, RU--94--67 (1994).

\bibitem{Grvz2}
J.-F. Grivaz, LAL--92--64 (1992).


\end{thebibliography}

\bigskip
\begin{center}
FIGURE CAPTIONS
\end{center}
\bigskip

\noindent
\small{Fig.~1. The characteristic regions of parameter space.}

\medskip
\noindent
\small{Fig.~2.  The allowed region of the $(\phi_+, \phi_-)$ plane.
Contours of $\mwchi$ are plotted in GeV.}

\medskip
\noindent
\small{Fig.~3. Allowed regions of the $(\msnu, \gchi)$ plane. Solid
(dotted) curves are $\sigma_L$ ($\afbtrunc$) constraints.}

\end{document}